\documentclass[9pt,twocolumn,twoside]{opticajnl}
\journal{opticajournal} 
\setboolean{shortarticle}{false}
\usepackage{lineno}
\usepackage{graphicx}
\usepackage{dcolumn}
\usepackage{amsmath,amsfonts,amssymb}
\usepackage{bm}
\usepackage{lipsum}
\usepackage[colorlinks=true, allcolors=blue]{hyperref}
\usepackage{url}
\usepackage{lineno}

\title{Polarization entanglement and qubit error rate dependence on the exciton-phonon coupling in self-assembled quantum dots}

\author[1,*]{Urmimala Dewan}
\author[2]{Parvendra Kumar}
\author[3]{Amarendra K. Sarma}

\affil[1]{Department of Physics, Indian Institute of Technology Guwahati, Guwahati- 781039, Assam, India}

\affil[2]{Optics  and  Photonics Centre, Indian Institute of Technology Delhi, Hauz Khas, New Delhi-110016, India}

\affil[3]{Department of Physics, Indian Institute of Technology Guwahati, Guwahati- 781039, Assam, India}

\affil[*]{d.urmimala@iitg.ac.in}
\setboolean{displaycopyright}{false}

\begin{abstract}
Polarization-entangled photons are key resources for a wide range of protocols in quantum computation and quantum key distribution. Achieving a near-unity degree of polarization entanglement is essential for minimizing qubit error rates in secure key distribution. In this work, we theoretically investigate polarization-entangled photon pairs generated via a quantum-dot radiative cascade embedded in a micropillar cavity. To account for the unavoidable exciton–phonon interactions in the quantum dot–cavity system, we develop a polaron master-equation framework and examine its impact on the degree of entanglement and the resulting qubit error rate. We derive analytical expressions for phonon-induced incoherent scattering rates and show that one-photon incoherent processes dominate, leading to a substantial reduction of entanglement. We further demonstrate that at elevated phonon-bath temperatures, cavity-mediated effects—such as cross-coupling between exciton states, ac-Stark shifts, and multiphoton emission—are significantly suppressed due to phonon-induced renormalization of the cavity coupling strength and the Rabi frequency. Finally, we analyze a BBM92 quantum key distribution protocol and study the evolution of the qubit error rate as a function of the phonon-bath temperature.
\end{abstract}

\begin{document}

\maketitle

\section{Introduction}\label{sec:level1}
In recent decades, entangled photon pairs have evolved from a purely theoretical concept into a key resource for practical applications in quantum communication \cite{Basset,Zopf}, quantum key distribution \cite{Yin} and quantum computing \cite{Brien}. In general, entanglement is categorized by the physical degree of freedom in which correlations between photons are realized. Prominent examples include polarized entangled photons \cite{nature_expt, Akopian}, time-bin entangled photons \cite{Christoph, Versteegh, Takesue}, hyperentangled photons \cite{Maximilian}, or frequency-bin entangled photons \cite{Kues}. Most state-of-the-art sources of entangled photon pairs are based on spontaneous parametric down-conversion processes \cite{Bernecker}. Owing to their probabilistic nature and the possibility of multipair emission, such sources are not ideally suited for applications in quantum key distribution and quantum computation. In contrast, radiative cascades in single quantum emitters, such as quantum dots (QDs), provide a deterministic alternative for generating entangled photon pairs \cite{Benson}. QDs are well \textcolor{blue}{established} as reliable on-demand sources of highly indistinguishable single photons \cite{Solomon,Bozzio} and entangled photons with near-unity quantum efficiency and compatibility with modern photonic chip integration \cite{Chen,AIP}. Significant progress has been made in generating entangled photon pairs with polarization entanglement from quantum dots \cite{Akopian,twinphotons}. Beyond polarization, quantum-dot–based sources have also enabled time-bin entanglement \cite{Jayakumar,THuber} and hyperentanglement as well \cite{Maximilian}. Moreover, polarization entanglement can be converted into time-bin entanglement, either probabilistically or using ultrafast optical modulators \cite{Aumann}. 

In a quantum dot, the biexciton decays radiatively via two intermediate optically active exciton states. A high degree of entanglement requires that the two decay paths have orthogonal polarizations while remaining otherwise indistinguishable. However, in-plane structural asymmetries of quantum dots lift the degeneracy of the exciton states, leading to an energy splitting known as the fine-structure splitting (FSS). The presence of FSS introduces \emph{which-path information} in the cascaded radiative decay, thereby limiting the achievable degree of entanglement \cite{Hudson, Zou, Varo}. Nevertheless, FSS can be mitigated by applying electric or strain fields or by growing QDs within highly symmetric structures such as nanowires \cite{Trotta,Juska}. 

Another inevitable source of decoherence in QDs is the longitudinal acoustic (LA) phonon coupling to the exciton and biexciton states \cite{Parvendra, Bester, Ramsay}. \textcolor{black}{Several theoretical models based on the quantum master equation, polaron and variational polaron transform \cite{Nazir,Gustin}, path integral and correlation expansion approaches \cite{Glassl,Barth,forstner} have been developed and used successfully to take into account the exciton-phonon coupling in different parameter regimes. While conventional quantum master equations are computationally efficient, they are generally valid only in the weak exciton–phonon coupling regime and at low temperatures \cite{Nazir}. In contrast, polaron and variational transform-based master equations give physically correct results beyond the weak exciton-phonon coupling and for a wide range of temperature and Rabi frequency \cite{Gustin, Nazir}, while providing deeper insight into the role of exciton-phonon coupling through the analytical form of various phonon-induced processes, which generally remain obscure in the complex numerical techniques based on path integral and correlation expansion approaches. Moreover, the master equation approaches allow the straightforward simulations of relevant quantities, including two-time correlation functions and fluorescence spectra \cite{Nazir, Kiraz}.}

It is worth noting that polarization entanglement in quantum dots has been extensively studied both in the absence \cite{Samal, Troiani} and presence of phonons \cite{twinphotons, Seidelmann, Kuhn, Cecoi}. However, the aforementioned works either consider an initially prepared biexciton state without including the pump pulse, which means that the exciton-phonon coupling mediated by the pump pulse is ignored, or investigate the emission of photons into free-space modes without including the cavity coupling and hence do not include the exciton-phonon coupling effects mediated by the cavity mode field. Note that both pump pulse and cavity coupling showed significant impact on the polarization entanglement \cite{Cosacchi, Schumacher}. In this paper, we develop a polaron transform-based quantum master equation (ME) that takes into account the exciton-phonon coupling associated with both pump pulse and cavity modes to investigate the polarization-entangled photon pairs. We compare the phonon-induced decay rates associated with the pump pulse and cavity modes that have been analytically calculated. The decay rates associated with the pump pulse are found to be significantly higher than the decay rates associated with the cavity modes. Furthermore, we show, among others, that phonon-mediated one-photon excitation and deexcitation processes play a dominating role and significantly degrade the degree of entanglement. Moreover, we demonstrate that at elevated phonon-bath temperatures, the cavity-mediated processes, including cross-coupling between exciton states, ac-Stark shift, and multiphoton emission, are diminished due to the phonon-induced renormalization of the cavity coupling and Rabi frequency. Finally, we show that the qubit error rate, which is the mismatch between Alice's and Bob's measurement outcomes, also increases at the increased temperatures.

This paper is organized as follows. Section \ref{sec2}, describes the theoretical description of the quantum dot cavity system, the computation of the concurrence from the two-photon density matrix, and the polaron master equation to incorporate the exciton-phonon interaction. In Sec. \ref{sec3}, we investigate and discuss the effects of exciton-phonon coupling on the degree of entanglement and qubit error rate by numerically solving the polaron master equation. Finally, we present our concluding remarks in Sec. \ref{sec4}. Appendices \ref{A} and \ref{B} provide insight into the details of the derivation of the polaron master equation.

\section{Theory}{\label{sec2}}

\subsection{Model and Two-photon density matrix}{\label{sec:level2}}

In this work, we consider a driven GaAs/InAs QD embedded in a micropillar cavity. To generate entangled photon pairs, we exploit the biexciton-exciton cascade in a quantum dot (QD) system. We model the QD as a four-level system composed of the ground state $\vert G\rangle$, horizontal exciton state 
$\vert H\rangle$, vertical exciton state $\vert V\rangle$, and biexciton state $\vert B\rangle$. In our model, the exciton and biexciton states are coupled to two degenerate horizontally and vertically polarized modes of a micropillar cavity, which facilitates the channeling of radiatively emitted photons into the corresponding polarized cavity modes. In a single QD, the decay of a biexciton state to ground state occurs via two equally probable pathways: one resulting in the emission of two horizontally polarized photons, and the other in the emission of two vertically polarized photons. The photons generated through the biexciton-exciton cascade ideally result in the formation of a polarization entangled two-photon state \cite{Seidelmann},
\begin{equation}
    \vert \Psi \rangle = \frac{(\vert HH\rangle + e^{i\delta t^{\prime}/\hslash}\vert VV \rangle )}{\sqrt{2}}
\end{equation}
here, $\delta$ is the fine structure splitting of exciton states and $ t^{\prime}$ represents the time delay between the biexciton and exciton photon emission events. In typical InAs/GaAs QDs, FSS lies in the range of a few tens of  $\mu$eV \cite{Seidelmann}. To prepare the QD biexciton state and collection of emitted photons into the orthogonally polarized degenerate cavity modes, we employ standard two-photon resonant excitation and emission processes, where the cavity frequency and laser frequency are set to satisfy the condition, $\omega_c = \omega_l = \omega_B/2$. Here, $\omega_c=\omega_{c_H} =\omega_{c_V}$ is the frequency of degenerate cavity modes, while $\omega_l$ and $\omega_B$ are frequencies of pumping laser and biexciton state, respectively.   We use concurrence to rigorously quantify the degree of entanglement \cite{Troiani}, which is obtained from the two-photon density matrix $\rho^{TP}$.
\begin{equation}
{\rho}^{TP} =
\begin{pmatrix}
\alpha_{HH} & \gamma_1 & \gamma_2 & \gamma \\
\gamma_1 & \beta_{HV} & \gamma_2 & \gamma_4 \\
\gamma_2 & \gamma_3 & \beta_{VH} & \gamma_5 \\
\gamma & \gamma_4 & \gamma_5 & \alpha_{VV} 
\end{pmatrix}
\label{Eq.2}
\end{equation}
The two-photon density matrix ${\rho}^{TP}$ characterizes the quantum state of photons emitted from the quantum dot. The matrix elements, $\alpha_{HH}$ and $\alpha_{VV}$ denote the probabilities of joint detection of co-polarized photon-pair emission, while, $\alpha_{HV}$ and $\alpha_{VH}$  quantify cross-polarized events. The off-diagonal elements quantify the polarization coherences between different two-photon states. These elements are reconstructed experimentally via polarization-resolved time-integrated correlation measurements in quantum state tomography \cite{White}. Theoretically, they are calculated using the quantum regression theorem applied to the corresponding correlation function, for example \cite{Zou,Troiani},
\begin{equation}
    \langle \mu \nu \vert \rho^{TP}\vert \xi\zeta\rangle = N \int_0^{T_p} dt \int_0^{{T_p}^{\prime}} dt^{\prime} \langle a^{\dagger}_\mu (t) a^{\dagger}_\nu(t+t^{\prime}) a_\zeta(t+t^{\prime}) a_\xi(t)\rangle 
\label{Eq.3}
\end{equation}
where ${T_p}$ and ${T_p}^{\prime}$ define the temporal window related to the detection of photons emitted from biexciton and exciton states, respectively. $N$ is the normalization constant and $\mu,\nu,\zeta,\xi \in H,V$.
Specifically, the diagonal elements are given by
$
\alpha_{HH} = \langle HH \vert \rho^{TP} \vert HH \rangle,$
$\alpha_{VV} = \langle VV \vert \rho^{TP} \vert VV \rangle,$ whereas the coherence term $\gamma$ is given by $\gamma = \langle HH \vert \rho^{TP} \vert VV \rangle.
$ Moreover, two photons in the entangled state can be distinguished experimentally by their different emission frequencies arising from the biexciton binding energy, which allows them to be spectrally separated into distinct modes using appropriate notch filters \cite{Schimpf}.
According to the Peres Criterion \cite{peres}, the polarization state exhibits entanglement if $\gamma\neq 0$, indicating that the \textit{which-path} information is erased due to the indistinguishability of the paths. The maximal entanglement arises at $\vert \gamma\vert = \frac{1}{2}$, indicating both paths are equally probable. However, in practice, non-zero values of $\beta_{HV}$, $\beta_{VH}$ result in degradation of entanglement. In such cases, the concurrence can be obtained directly from the two-photon density matrix $\rho^{TP}$ by calculating the four eigenvalues $e_{j}$ of the matrix $M$ = $\rho^{TP}\mathcal{A}(\rho^{TP})^{*}\mathcal{A}$,  where $(\rho^{TP})^{*}$ represents the complex conjugated two-photon density matrix, and $\mathcal{A}$ is the anti-diagonal matrix with elements (-1,   1,   1,-1). The concurrence $C$ is defined as, $C$ = $max(0,  \sqrt e_{1}-\sqrt e_{2}-\sqrt e_{3}-\sqrt e_{4})$, eigenvalues are sorted in decreasing order $e_{j+1}\leq e_{j}$ \cite {Troiani}. 

The qubit error rate (QBER), arising from imperfections in entangled state generation due to various decoherence processes, is calculated as the erroneous coincidence counts over the total number of detections within a given time window. The QBER is obtained directly from the two-photon density matrix \cite{Schimpf, Gisin},
\begin{equation}
    q = \frac{1}{2}\sum_{i = 1}^{4}\langle O_i\vert \rho^{TP}\vert O_i\rangle
\end{equation}
where $O_i\in \{HV,VH,DA,AD\}$, measured in the orthogonal basis, corresponds to the correlation between the emitted photons.

\subsection{Hamiltonian and Polaron master equation}{\label{sec:level2}}

\begin{figure*}[htbp]
    \centering
    \includegraphics[width=.85\linewidth]{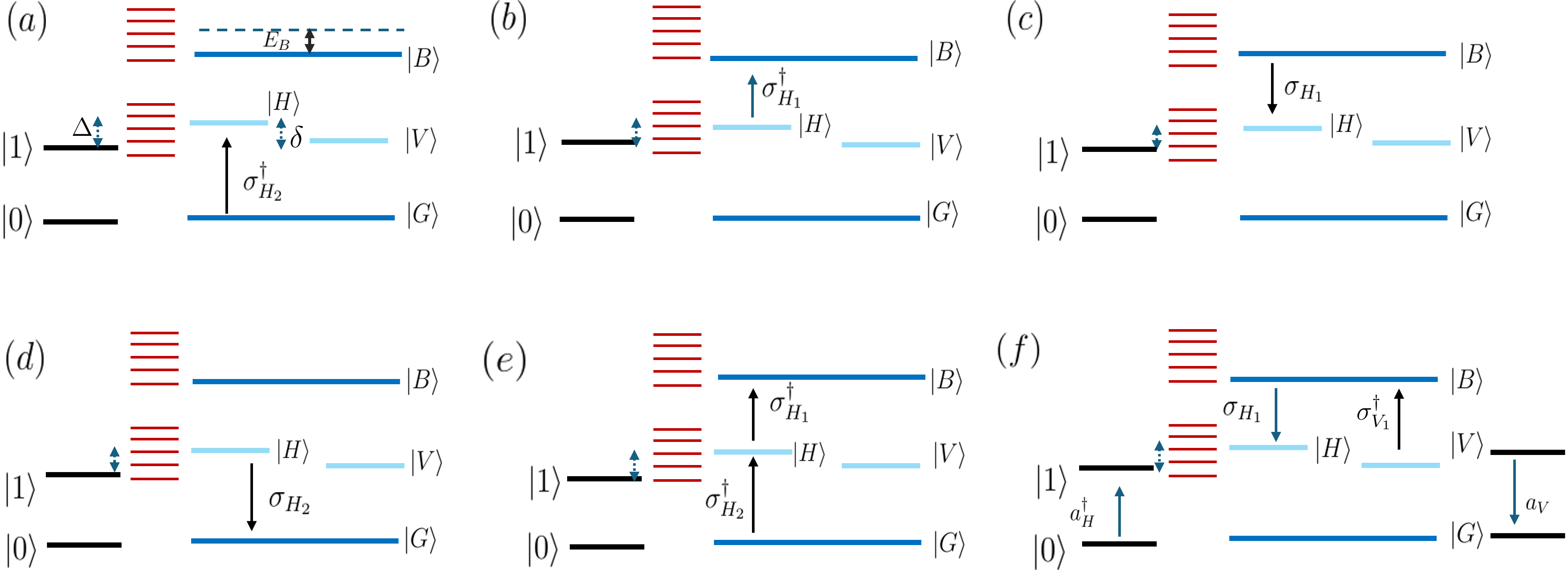}
    \caption{(color online) Schematic diagram of various phonon-mediated incoherent excitation and de-excitation. (a) and (b) phonon-induced one-photon incoherent excitation of H-polarized exciton and biexciton states, (c) and (d) phonon-induced one-photon de-excitation of $H$ polarized exciton and biexciton states, (e) phonon-assisted two-photon incoherent excitation of the biexciton state, and (f) phonon-induced cross-coupling between the $H$ and $V$ polarized exciton states. }
    \label{f1}
\end{figure*}

We consider that a horizontally polarized laser pulse and cavity modes interact with the QD, satisfying the two-photon resonant excitation of the biexciton state. Furthermore, we consider spatially orthogonal excitation of the quantum dot and collection of cavity emission to reduce the laser scattering into the cavity modes. However, resonance between the pump pulse and the cavity modes can, in principle, lead to pump-photon leakage and reduce the measured concurrence; this effect is effectively suppressed by standard temporal gating of photon-pair detection after the excitation pulse \cite{Young}. By opening the detection window only once the pump has ended, residual pump photons decay before significant biexciton-cascade emission into the cavity modes, ensuring that pump leakage does not contribute appreciably to the measured two-photon correlations; while temporal gating may reduce brightness, the concurrence remains largely unaffected. The Hamiltonian of the QD-cavity system under the rotating wave approximation can be written as \cite{Samal},
\begin{align}
    &H = H_{QD}+H_{QD-cav}+H_H 
    \\
    &H_{QD} = \frac{\hslash}{2}(E_B/ \hslash +\delta)|H\rangle \langle H| + 
    \frac{\hslash}{2}(E_B/ \hslash -\delta)|V\rangle \langle V| \\ 
    &H_{QD-cav} = \hslash g ( a^{\dagger}_H |G\rangle \langle H| + a^{\dagger}_H |H\rangle \langle {B}| 
    + a^{\dagger}_V |G\rangle \langle V| \\ \notag
    &+ a^{\dagger}_V |V \rangle \langle {B}|+ H.c.)
    \\ 
    &H_H = \frac{\hslash \Omega_H(t)}{2} ( |G\rangle \langle H| + |H\rangle \langle {B}|+ H.c.)
    \\
    &H_B   =\hslash \sum_q \omega_q b^{\dagger}_q b_q
    \\
    &H_I  = (\vert H \rangle \langle H \vert +\vert V \rangle \langle V\vert+ 2\vert B \rangle \langle B \vert  ) \sum_q \hslash \lambda_q (b_q + b^{\dagger}_q),
\end{align}
here, $H_{QD}$ is the free-energy Hamiltonian of QD, $H_{QD-cav}$ represents the coupling of horizontally ($H$) and vertically ($V$) polarized cavity modes to the identically polarized exciton and biexciton states. We assume that the orthogonally polarized cavity modes are degenerate and their coupling to exciton and biexciton states are identical, i.e., $g_H$ = $g_V$ = $g$. Moreover, $\Omega_{H} (t)$ represents the coupling of the horizontally polarized pulse laser with the QD, $a^\dagger_{H(V)}$ and $a_{H(V)}$ are creation and annihilation operators corresponding to the $H$($V$) polarized photons. The Rabi frequency associated with the horizontal pulse is defined as, $\Omega_H(t) = \Omega_{H_0}  e^{-(t-t_0)^2/t_p^2}$, where $\Omega_{H_0} = \frac{E_0 d}{\hslash}$. Here, $E_0$ is the electric field amplitude of the laser drive and $d$ represents the transition dipole moment of the exciton and biexciton states, while $E_B =  2\hbar(\omega_H - \delta/2) -\hbar \omega_B, $ is the biexciton binding energy, $\omega_H$ horizontally polarized exciton frequency. 
We have considered that each energy level of the QD is coupled to a phonon bath, which is treated as a collection of harmonic oscillators.
The phonon annihilation (creation) operator of the $q$th mode is given by $b^{(\dagger)}_q$ with frequency $\omega_q$. The longitudinal acoustic phonon-exciton coupling is included via coupling constants $\lambda^{s}_{\textbf{q}}$ for $s = \{H,V,B\}$, which correspond to an ideal quantum confined QD such that $\lambda_{\textbf{q}}$ = $\lambda^{H}_{\textbf{q}}$ = $\lambda^{V}_{\textbf{q}}$ = $\frac{1}{2} \lambda^{B}_{\textbf{q}}$ \cite{Hohenester}.
As a biexciton state $\vert B\rangle$ consists of two independent excitons with opposite spin orientations, therefore its coupling to phonon bath is approximately twice that of an excitonic state.
A schematic representation of the model is illustrated in Fig. \ref{f1}(a). The energy levels of quantum dots are represented by blue lines, the phonon bath associated with quantum dot states is indicated by red lines, and the energy levels of cavity modes are depicted by black lines.
To treat exciton-phonon coupling non-perturbatively, we carry out a unitary polaron transformation,
$H^{\prime} = e^P H e^{-P}$,
with $P = \left ( \vert V \rangle \langle V \vert + \vert H \rangle \langle H \vert + 2\vert B \rangle \langle B \vert \right )\sum_q \frac{\lambda_q}{\omega_q}(b^{\dagger}_q-b_q)$, to diagonalize the exciton-phonon coupling part of the Hamiltonian \cite{Gustin,Hazra}. The polaron frame transformed Hamiltonian, $H^{\prime} = H_S^{\prime}+ H_B^{\prime}+H_I^{\prime}$ would be,
\begin{align}
   &H_S^{\prime} = \hslash \Delta \vert {H}\rangle  \langle {H}\vert + 
     \hslash (\Delta-\delta) \vert {V}\rangle  \langle {V}\vert +\langle \mathcal{B} \rangle X_g(t)
   \\
   & H_B^{\prime} =\hslash \sum_q \omega_q b^{\dagger}_q b_q  \label{appb}
   \\
  &H_I^{\prime} = X_g \zeta_g +X_u \zeta_u \label{appc},
\end{align}
here, $\Delta = (E_B/\hbar+\delta)/2 = (\omega_H- \omega_l), $ is the detuning between the $H$ polarized exciton state and pump laser.
The polaron transformed system operators are given by,
\begin{align}
    &X_g(t) = \hslash g (a^{\dagger}_H \vert G \rangle \langle H \vert +a^{\dagger}_H \vert H \rangle \langle B \vert + a^{\dagger}_V \vert G \rangle \langle V \vert  \notag \\ \notag
    &+a^{\dagger}_V \vert V \rangle \langle B \vert) + \frac{ \hslash \Omega_H(t)}{2}(\vert G \rangle \langle H \vert + \vert H \rangle \langle B \vert ) +H.c \\ \notag
    &X_u(t) =i \hslash g  (a^{\dagger}_H \vert G \rangle \langle H \vert +a^{\dagger}_H \vert H \rangle \langle B \vert  + a^{\dagger}_V \vert G \rangle \langle V \vert \\ \notag
    &+a^{\dagger}_V \vert V \rangle \langle B \vert) 
    + \frac{i\hslash \Omega_H(t)}{2}    
    (\vert G \rangle \langle H \vert + \vert H \rangle \langle B \vert)+H.c
\end{align}
The operators, $\zeta_g = \frac{1}{2}(\mathcal{B}_+ +\mathcal{B}_- - 2\langle \mathcal{B} \rangle)$, and $\zeta_u= \frac{1}{2i}(\mathcal{B}_+ + \mathcal{B}_-)$, represent the phonon-bath fluctuations, while $\mathcal{B}_{\pm} = \exp{[\pm \sum_q \frac{\lambda_q}{\omega_q} (b^{\dagger}_q -b_q)]}$ are the displacement operators with equal expectation value, $\langle \mathcal{B} \rangle = \langle \mathcal{B}_+ \rangle = \langle \mathcal{B}_- \rangle = \exp[{-\frac{1}{2}\sum_q \frac{\lambda^2_q}{\omega^2_q} \coth(\frac{\hslash \omega_q}{2k_B T})] }$.
The renormalization constant $\langle \mathcal{B} \rangle$ reduces the coupling strength of pumping laser pulse to $\langle \mathcal{B} \rangle \Omega_H(t)$ and cavity modes $\langle \mathcal{B} \rangle g$ with the QD states at increased temperatures.
In the continuum limit of phonon modes, we can characterize the electron-phonon interaction with the phonon spectral density function $J_p( \omega) = \sum_q \lambda^2_q \delta (\omega - \omega_q) \rightarrow  J_p(\omega) = \alpha_p \omega^3 \exp[\frac{-\omega^2}{2\omega^2_b}]$, that describes a deformation potential
induced by LA phonons, here $\alpha_p$ and $\omega_b$ are the electron-phonon coupling strength and the phonon cut-off frequency. The strength of the electron-phonon coupling depends on the composition of the surrounding host as well as the quantum dot material. The parameter $\alpha_p$ can be characterized in terms of deformation potential coupling constants $D_c$ and $D_v$ for the conduction and valence bands, as $\alpha_p = (D_c-D_v)^2/4\pi^2c_s^5d_{\rho}$, $c_s$ is the velocity of longitudinal sound waves and $d_{\rho}$ is the density of the material.

Following the procedural details pulse driven \cite{Nazir,Manson,Hughes,Carmicheal}, we derive a time local polaron master equation using second-order Born-Markov approximation, given as,
\begin{equation}
    \begin{split}
        \frac{d\rho}{dt} =  \frac{1}{i\hbar}[ H^{\prime}_S(t),\rho(t)] +\mathcal{L}_{cav}\rho+\mathcal{L}_{rad}\rho+\mathcal{L}_{deph} \rho + \mathcal{L}_{ph}\rho
    \label{Eq.8}
    \end{split}
\end{equation}
Lindblad polaron dissipator $\mathcal{L}_{ph}\rho$ is defined as,
\begin{align}
    \mathcal{L}_{ph}\rho = -\frac{1}{\hslash^2}\int_{0}^{\infty} \sum_{m= g,u} d\tau
     \{ G_m(\tau) [X_m(t),X_m (t,\tau)\rho(t)] 
    +H.c. \}. 
    \label{Eq.9}
\end{align}

Here, $X_m(t,\tau ) = e^{-iH^{\prime}_S\tau/ \hslash}X_m e^{iH^{\prime}_S\tau/\hslash} $, 
$\mathcal{L}[\hat{O}] = 2\hat{O}\rho \hat{O}^\dagger - \hat{O}^\dagger\hat{O}\rho- \rho \hat{O}^\dagger \hat{O}$ and  $\mathcal{L}[\hat{a},\hat{b}] = 2\hat{a}\rho\hat{b}^\dagger -\hat{a}^\dagger\hat{b}\rho- \rho \hat{a}^\dagger \hat{b}$. $G_g(\tau) =\langle \mathcal{B}\rangle^2 \{ \cosh[\phi (\tau)] - 1\} $ and $G_u(\tau) = \langle \mathcal{B}\rangle^2 \sinh{[\phi(\tau)]}$ are the polaron Green's functions, and $\phi(\tau) = \int^{\infty}_0 d\omega \frac{J_p(\omega)}{\omega^2}[\coth(\frac{\hslash \omega}{2k_B T}) \cos{(\omega \tau)}-i\sin{(\omega\tau)}]$ is the phonon correlation function at temperature of phonon-bath ${T}$. For convenience, we have denoted the QD operators as follows,
$\sigma_{H_1} = \vert H \rangle \langle B \vert $, $\sigma_{H_2} = \vert G \rangle \langle H \vert$,
$\sigma_{V_1} = \vert V \rangle \langle B \vert$ and $\sigma_{V_2} = \vert G \rangle \langle V \vert$.
The Liouvillian terms are defined as, $\mathcal{L}_{rad}\rho = \frac{\gamma_B}{2}[L(\sigma_{H_1})\rho+L(\sigma_{V_1})\rho]+\frac{\gamma_E}{2}[L(\sigma_{H_2})\rho+L(\sigma_{V_2})\rho] $ is the radiative relaxation of biexciton and exciton states, $\mathcal{L}_{deph}\rho =\frac{\gamma^{\prime}_B}{2} [L(\vert B\rangle\langle B\vert)\rho]+ \frac{\gamma^{\prime}_E}{2} [L(\vert H\rangle\langle H\vert)\rho+ L(\vert V\rangle\langle V\vert)\rho]$ represents the dephasing of the biexciton and exciton states, and cavity decay is incorporated via $\mathcal{L}_{cav}\rho = \frac{\kappa}{2} [L(a_H)\rho+L(a_V)\rho]$.
$\gamma_B,\gamma_E$ are the radiative decay rates of exciton and biexciton rates, $\gamma^{\prime}_B, \gamma^{\prime}_E$ are the dephasing rates of biexciton and exciton. $\kappa $ represents the cavity decay rate of $H$ polarized and $V$ polarized photons. It is worth noting that we treat the quantum dot exciton states together with the cavity modes as the system, while phonons are modeled as a thermal bath. The coupling between the exciton states and lattice vibrations is described microscopically via the dominant deformation-potential interaction. In addition, the coupling of both the cavity modes and the exciton states to the photonic reservoir is included phenomenologically, assuming weak coupling to the photonic bath \cite{Hughes,Manson, hughescarmi}.

For clarification of the role of exciton-phonon coupling, we compute the commutator in Eq. \ref{Eq.9} by approximating $X_m(t,\tau ) = e^{-iH^{\prime}_S\tau/\hslash}X_m e^{iH^{\prime}_S\tau/\hslash} \approx e^{-iH^{\prime}_0\tau/\hslash}X_m e^{iH^{\prime}_0\tau/\hslash}$, where $H^{\prime}_S$ is full polaron transformed Hamitonian and $H_0^{\prime}= \hslash \Delta \vert {H}\rangle  \langle {H}\vert + 
 \hslash \Delta \vert {V}\rangle  \langle {V}\vert +  \frac{\hslash \Omega^{\prime}_H(t)}{2}(\sigma^{\dagger}_{H_1}+\sigma^{\dagger}_{H_2}+ \sigma_{H_1}+\sigma_{H_2})$, where, $ \Omega^{\prime}_H(t) = \langle \mathcal{B} \rangle \Omega_H(t)$. This approximation corresponds to neglecting the quantum-dot–cavity interaction in the time-evolved polaron operator $X_m(t,\tau )$, while keeping all other terms unchanged. It significantly simplifies the evaluation of the phonon correlation functions and is justified if QD–cavity coupling is much smaller than the detuning $\Delta$ . In the present work, the maximum coupling strength $g$ is $70\mu eV$, which is much smaller than the detuning $\Delta = 1.1 meV$, thereby validating the approximation. The analytical expressions of various phonon-induced incoherent rates are as given in Appendix \ref{A} and \ref{B}. Below, we present a few phonon-induced rates. We find $\Gamma^{\pm}_{\Omega}$ to be at least one  order of magnitude larger than the other phonon-induced rates and therefore contributes significantly to the overall system dynamics.
\begin{align}
    &\Gamma^{\pm}_{\Omega}=  (\frac{\Omega_H(t)}{2} )^2 \langle \mathcal{B} \rangle^2 \int_{0}^{\infty} d\tau Re \left \{ (e^{\phi(\tau)}-1) e^{\pm i \Delta \tau} \right \} 
    \label{Eq.12}\\
    &\Gamma^{\pm}_H= g^2 \langle \mathcal{B} \rangle^2 \int_{0}^{\infty} d\tau Re \left \{ (e^{\phi(\tau)}-1) e^{\pm i \Delta \tau} \right \}
    \label{Eq.13} \\
     &\Gamma^{\pm}_V= g^2 \langle \mathcal{B} \rangle^2 \int_{0}^{\infty} d\tau Re \left \{ (e^{\phi(\tau)}-1) e^{\pm i (\Delta-\delta) \tau} \right \}
    \label{Eq.14} \\
    &\Gamma^{TP}_{\Omega}=  (\frac{\Omega_H(t)}{2} )^2 \langle \mathcal{B} \rangle^2 \int_{0}^{\infty} d\tau Re \left \{ (e^{-\phi(\tau)}-1) e^{- i \Delta \tau} \right \} \label{15}
\end{align}

It is clear from Eq. \ref{A1} in Appendix \ref{A} that $\Gamma^+_{\Omega}$ describes the phonon-assisted incoherent excitation (de-excitation) of the biexciton (exciton) states as shown in Fig. \ref{f1}(b) (\ref{f1}(d)), while $\Gamma^-_{\Omega}$ represents the phonon-assisted incoherent excitation (de-excitation) of exciton (biexciton) states in the presence of the laser drive as depicted in Fig. \ref{f1}(a) (\ref{f1}(c)). Similarly, $\Gamma^{+}_H$ and $\Gamma^{+}_V$ correspond to the excitation (de-excitation) of the biexciton (exciton) state via cavity photon absorption (emission), $\Gamma^{-}_H$ and $\Gamma^{-}_V$ represents the de-excitation (excitation) of the biexciton (exciton) state via cavity photon emission (absorption) and $\Gamma^{TP}_{\Omega}$ represents the phonon-mediated two-photon excitation, shown in Fig. \ref{f1}(e), and de-excitation of the biexciton state associated with the pump pulse, whereas Fig. \ref{f1}(f) depicts phonon-mediated cross-coupling between the orthogonally polarized excitonic states due to cavity effects.

\section{Results and discussions}{\label{sec3}}
\begin{figure}[htbp]
    \centering
    \includegraphics[width=.82\linewidth]{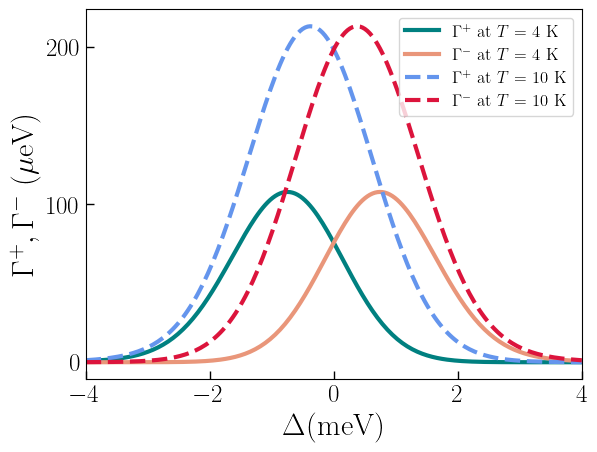}
    \caption{(color online) Phonon-mediated scattering rates  with \( \delta\) = 20 \(\mu\)eV and $ g$ = 70 $\mu eV $, (a) $\Gamma^{+}$ and $\Gamma^{-}$ at \( T = 4 \, \mathrm{K} \) and \( T = 20 \, \mathrm{K} \).(b) \( \langle \Gamma^{+}_{\Omega} \rangle\) and  $\langle \Gamma^{-}_{\Omega} \rangle$ at \( T = 4 \, \mathrm{K} \) and \( T = 20 \, \mathrm{K} \). $\langle \Gamma^{\pm}_\Omega \rangle$ are the time averaged decay rates over one pulse width, given as, $\langle \Gamma^{\pm}_\Omega \rangle$ $=$ $\int_{-\infty}^{\infty}  \Gamma^{\pm}_\Omega$ $dt /2t_p$. }
    \label{2}
\end{figure}
\begin{figure}[htbp]
    \centering
    \includegraphics[width=.84\linewidth]{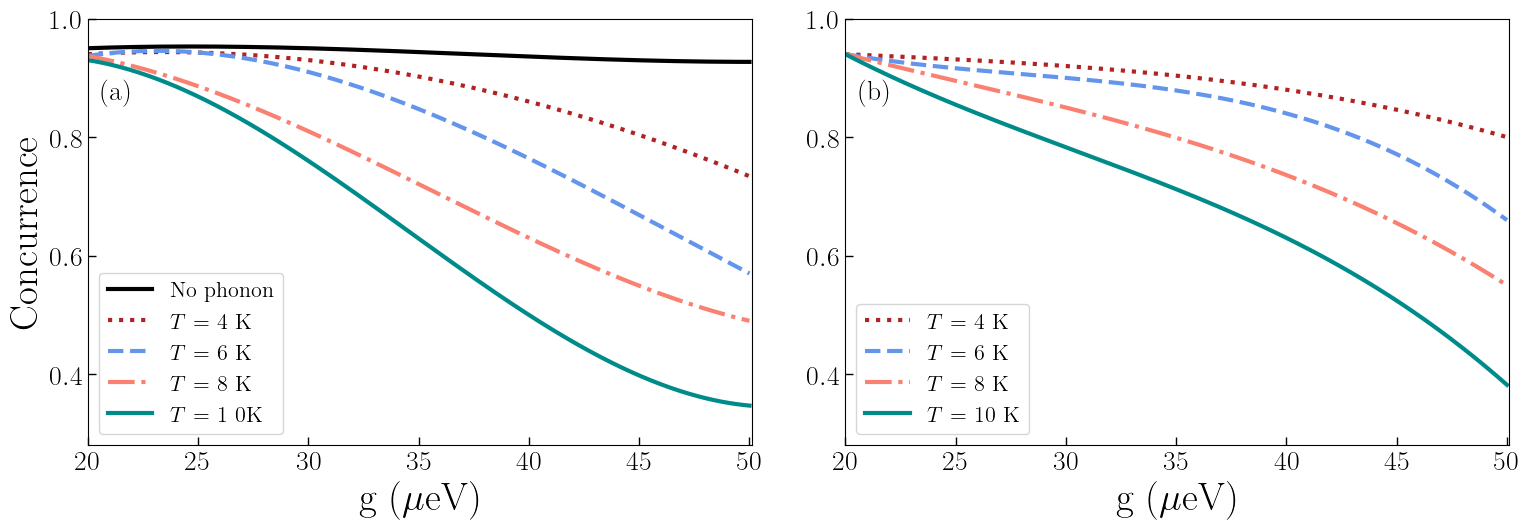}
    \caption{(color online) Different Phonon-mediated two-photon scattering rates  with \( \delta \) = 20  \(\mu\)eV and $ g$ = 70  $\mu eV $, (a) $\Gamma^{TP}$  at \( T = 4 \, \mathrm{K} \) and \( T = 20 \, \mathrm{K} \).(b) \( \langle \Gamma^{TP}_{\Omega} \rangle\) at \( T = 4 \, \mathrm{K} \) and \( T = 20 \, \mathrm{K} \). $\langle \Gamma^{TP}_\Omega \rangle$ is the time-averaged decay rate over one pulse width, given as, $\langle \Gamma^{TP}_\Omega \rangle$ $=$ $\int_{-\infty}^{\infty}  \Gamma^{TP}_\Omega$ $dt /2t_p$.} 
    \label{f3}
\end{figure}
\begin{figure*}[htbp]
    \centering
    \includegraphics[width=.75\linewidth]{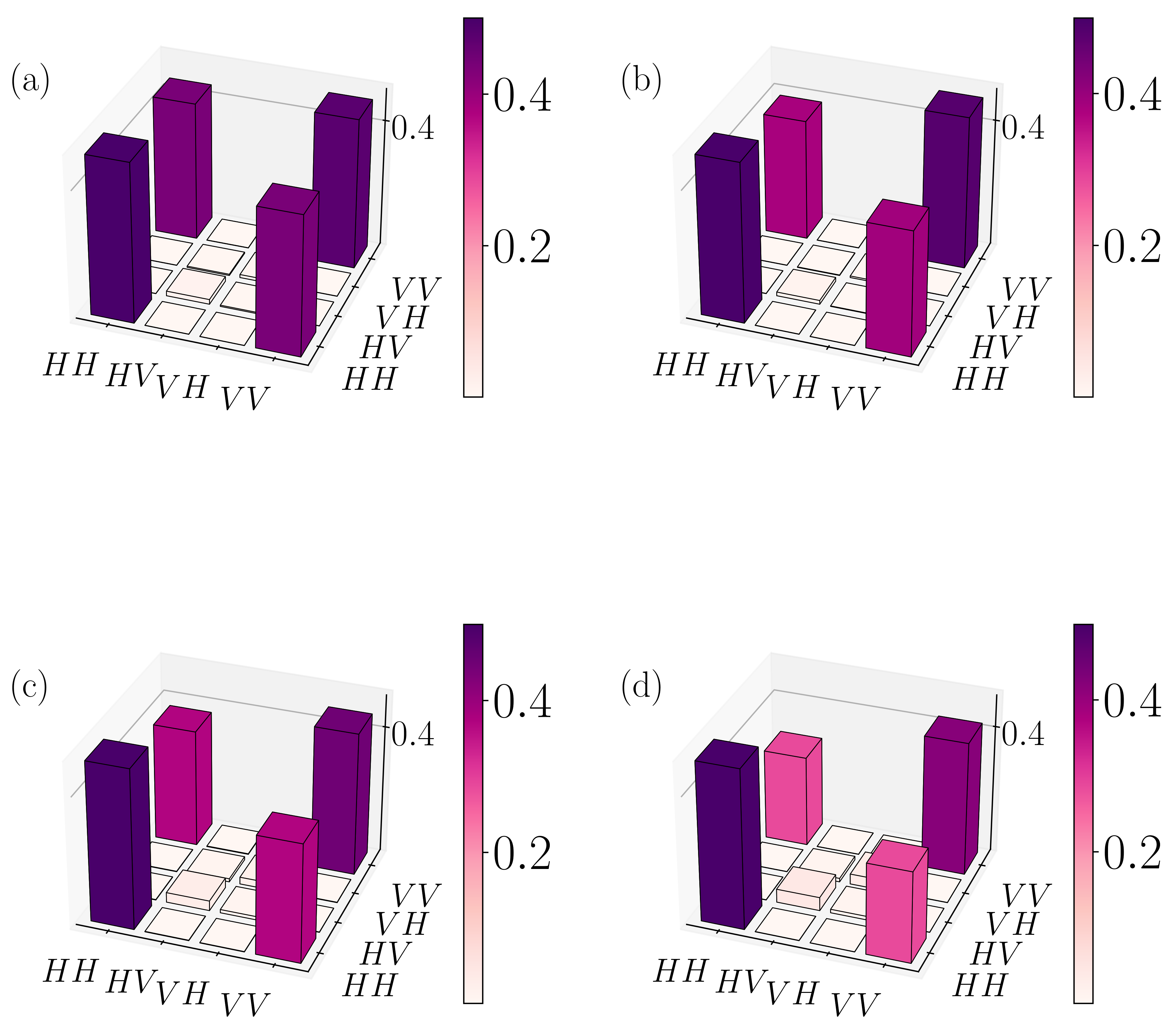}
    \caption{(color online) Concurrence as a function of $g$ for different temperatures with (a) $ {T_p}^{\prime} = 200$ ps and $\delta = 0$ (b) ${T_p}^{\prime} = 50$ ps and $\delta = 0$, (c) ${T_p}^{\prime}= 200$ ps, for $\delta = $ 0, 20 $\mu$eV (d) ${T_p}^{\prime} = 50$ ps, for $\delta = $ 0, 20 $\mu$eV.}
    \label{3}
\end{figure*} 

To investigate the role of exciton-phonon coupling on the entanglement and qubit error rate, we adopt the typical parameters for GaAs/InAs quantum dots (QDs) \cite{Hughes,hughescarmi}.
These include \( \alpha_p = 0.06\) \( ps^2 \), $\Delta = 1.1$ meV, \(\omega_b =  \)1 meV , $\Omega_{H_0} = 0.8$ meV, \(\gamma_B = 2 \) \(\mu  \)eV,  \(\gamma_E = 1 \) \(\mu \)eV,  \(\gamma^{\prime}_B = 4 \) \(\mu  \)eV, $t_p = 6$ $ps$, \(\gamma^{\prime}_E = 2\) \(\mu \)eV and \(\kappa = 65 \) \( \mu \)eV, and $\delta = 20$ $\mu$eV. For the chosen parameters throughout this work, the validity condition, $(\Omega_{H_0}/\omega_b)^2 (1-\langle \mathcal{B}\rangle^4)\ll1$, of the polaron master equation is satisfied \cite{Nazir,McCutcheon2010,Manson}.

From the numerical computation, we find that $\Gamma^+_H $ $\approx$  $\Gamma^+_V$ and $\Gamma^-_H $ $\approx$  $\Gamma^-_V$ and in the following discussion hereafter we denote them as $\Gamma^+_H $ $\approx$  $\Gamma^+_V$ $ = $ $\Gamma^+$, $\Gamma^-_H $ $\approx$  $\Gamma^-_V$ $ = $ $\Gamma^-$.
In Fig. \ref{2}, we illustrate the variation of $\Gamma^+$, $\Gamma^-$, and \(\Gamma^{\pm}_{\Omega}\) as functions of detuning \(\Delta\). As shown in Fig. \ref{2}, phonon-assisted processes are asymmetric at lower temperatures. The phonon emission rates are higher than the phonon absorption rates because at lower temperatures, the phonon absorption from the phonon bath is strongly reduced.
This primarily means that, at lower temperatures, for $\Delta > 0 $, the transfer of energy from the laser drive to the biexciton state along with emission of a phonon at the rate of $\Gamma^{+}_{\Omega},$ is preferred over the transfer of energy from the biexcition state to the laser field along with absorption of a phonon at a rate of $\Gamma^{-}_{\Omega}$. As expected, phonon-assisted scattering rates increase with temperature, showing a more symmetric nature.
This directly impacts the population of exciton and biexciton states.
In Fig. \ref{f3}, we show the phonon-mediated two-photon incoherent rates, $\Gamma^{TP}_{H}$ $ \approx$ $\Gamma^{TP}_{V}$ $\approx$ $\Gamma^{TP}$, associated with cavity coupling and phonon-induced two-photon incoherent rates, $\Gamma^{TP}_{\Omega}$, due to pump pulse (see Appendix \ref{A}). It is clear from Fig. \ref{2} and Fig. \ref{f3} that the phonon-mediated incoherent processes associated with pump pulse dominate over their cavity coupling counterparts. Overall, phonon-mediated one-photon incoherent rates associated with pump pulse as shown in Fig. \ref{2} (a) dominate over all the other rates.

\begin{figure}[h!]
    \centering
    \includegraphics[width=1 \linewidth]{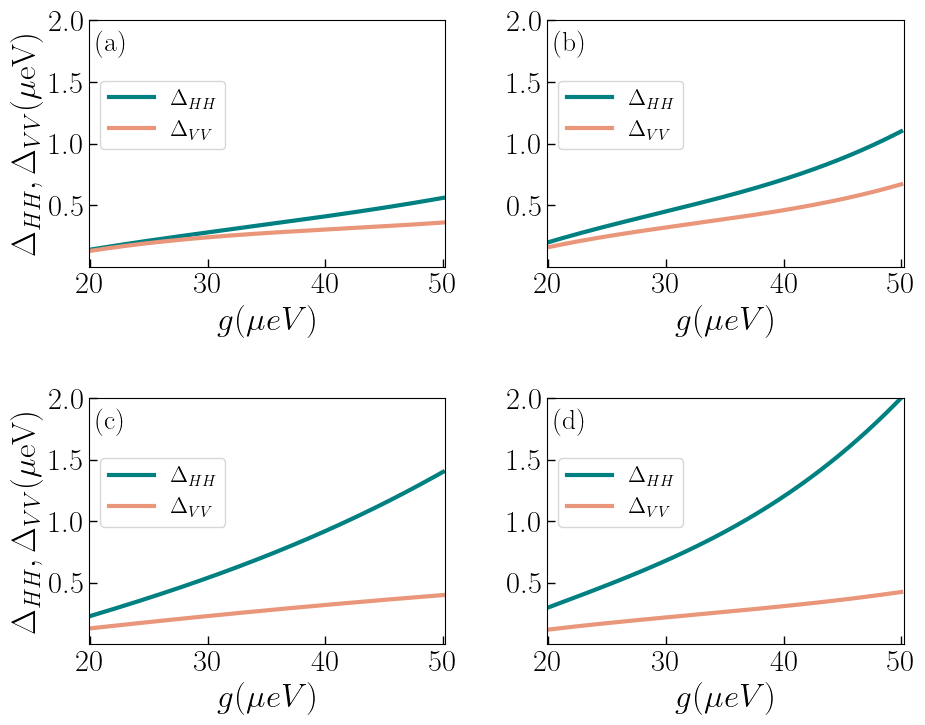}
    \caption{(color online) Two-photon density matrix of polarization-entangled state at $g$ = 70 $\mu$eV for (a) \( T = 4 \, \mathrm{K} \), (b) \( T = 8 \, \mathrm{K} \) (c) \( T = 16 \, \mathrm{K} \) and (d) \( T = 20 \, \mathrm{K} \). As the temperature rises, a dip in the off-diagonal elements is visible due to reduced coherence between the states.}
    \label{4}
\end{figure}

As a measure of the degree of entanglement, we investigate the concurrence by numerically computing the two-photon density matrix $\rho^{TP}$ with the aid of Eq. \ref{Eq.3} and Eq. \ref{Eq.8}. In Fig. \ref{3}, we illustrate the concurrence as a function of the coupling strength $g$ at various temperatures for two time windows, ${T_p}^{\prime} = $ 200 ps and 50 ps. Without phonon coupling, as shown in Fig. \ref{3}(a), a maximum value of concurrence of 0.91 can be obtained at the lowest coupling strength, $g$ = 30 $\mu eV$ for $\delta = 0$. However, with phonon coupling, the concurrence decreases significantly at the elevated temperatures of the phonon bath. This occurs due to the partial loss of two-photon coherence, as shown in Fig. \ref{4}. Nevertheless, as evident from Fig. \ref{2}(b), the phonon-induced incoherent excitation and de-excitation processes associated with the pumping pulse dominate the cavity-mediated phonon decay rates shown in Fig. \ref{2}(b). As a consequence, the concurrence shows a mild dependence on the cavity coupling strength $g$ and cavity mediated phonon decay rates. Fig. \ref{3}(b) indicates that applying temporal filtering can effectively improve concurrence, particularly at elevated temperatures,  by partially eliminating the decoherence due to the time-dependent phonon incoherent rates. Fig. \ref{3}(c) and (d) show the effects of fine structure splitting on the degree of entanglement. In both cases of ${T_p}^{\prime}  = 200 $ ps and ${T_p}^{\prime}  = 50$ ps, there is a visible drop in concurrence even at $\delta$ = 20 $\mu$eV, particularly at $T$ = 20 K. Moreover, we separately find that the phonon-mediated two-photon excitation and de-excitation of the biexciton state $\Gamma_\Omega^{\textit{TP}}$ [see Appendix $\ref{A}$] does not reduce the concurrence appreciably even at increased coupling strength and higher temperatures.

In Fig. \ref{4}, we show the two-photon density matrix for illustrating the effects of phonon-induced incoherent rates on the coherence between two-photon states at different temperatures for $g$ = 70 $\mu$eV. The off-diagonal density matrix elements \(\rho_{HH,VV}\) and \(\rho_{VV,HH}\), represent the coherence between the \(|HH\rangle\) and \(|VV\rangle\) states, which is essential for the realization of a polarization-entangled two-photon state. With increasing temperature, a noticeable reduction in coherence is observed. Moreover, a non-zero value of two-photon states with orthogonal polarization \(|HV\rangle\) in Fig. \ref{4}(a) reflects the cavity-mediated cross-coupling between exciton states \cite{Samal}, whose magnitude is reduced at the increased temperatures due to the phonon-induced renormalization of $g$  by a factor of $\langle \mathcal{B} \rangle$. Additionally, we notice the decay dynamics show a preference for the $H$ polarized exciton channel over the $V$ polarized exciton, leading to a slight increase in \(\rho_{HH,HH}\) over \(\rho_{VV,VV}\) at elevated temperatures. This occurs because of the excitation of the QD by a horizontally polarized laser pulse.

\begin{figure}[ht]
    \centering
    \includegraphics[width=1
    \linewidth]{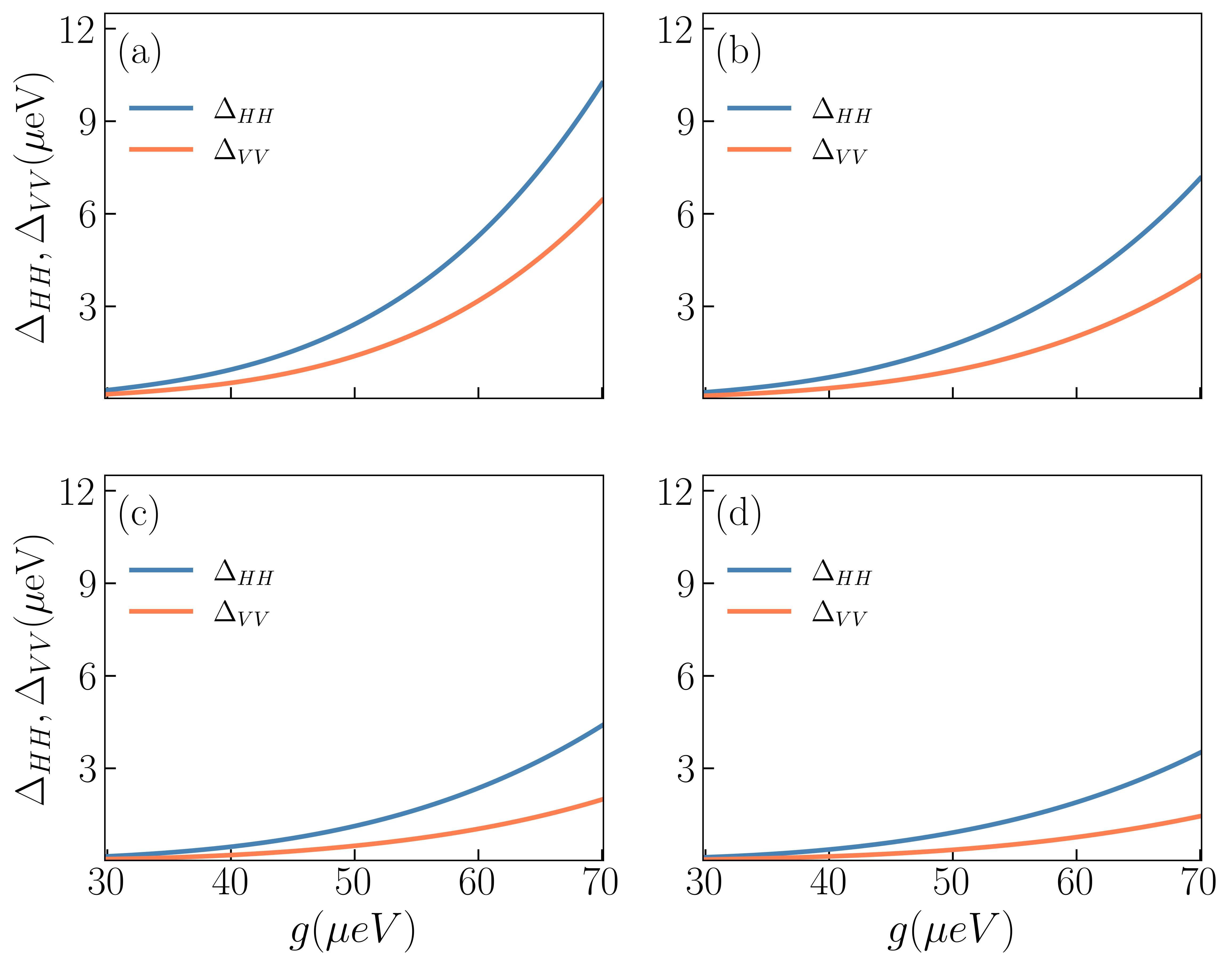}
    \caption{(color online) ac-Stark shift in H and V polarized excitons as a function of $g$ at (a) \( T = 4 \, \mathrm{K} \) (b) \( T = 8 \, \mathrm{K} \), (c) \( T = 16 \, \mathrm{K} \) and (d) \( T = 20 \, \mathrm{K} \).}
    \label{5}
\end{figure}

In addition to FSS, another potential cause of reduced coherence could be the cavity-field-induced ac-Stark shifts, which introduce the effective splitting of horizontally and vertically polarized exciton states. These are given as $\Delta_{HH} = 2\langle a^{\dagger}_H a_H\rangle \frac{({\langle \mathcal{B}\rangle g})^2}{\delta_H}$ and $\Delta_{VV} = 2\langle a^{\dagger}_V a_V\rangle \frac{(\langle \mathcal{B} \rangle g)^2}{\delta_V}$ for the $H$ and $V$ polarized exciton states, respectively \cite{starkshift}, $\delta_H =(E_B/\hslash+\delta)/2$ and $\delta_V =(E_B/\hslash-\delta)/2$. Here, $\langle a^{\dagger}_H a_H\rangle$, $ \langle a^{\dagger}_V a_V\rangle$ are the average photon numbers of $H$ and $V$ polarized photons. As depicted in Fig. \ref{5}, these energy shifts are unequal for the $H$ and $V$ polarized excitons, resulting in an effective energy splitting between the two exciton states.

This occurs as the $H$ polarized laser pulse interacts only with the $\vert H\rangle$ state, facilitating increased photon emission into $H$ polarized mode, resulting in a higher value of $\langle a^{\dagger}_H a_H\rangle$ as compared to $\langle a^{\dagger}_V a_V\rangle$. Moreover, as the temperature rises, the energy splitting gradually decreases due to the phonon-mediated renormalization of the cavity coupling strength $g$. This splitting, along with the FSS, introduces \textit{which-path} information, contributing to a further reduction in concurrence. At higher temperatures, the biexciton decays through the excitonic states instead of directly decaying to the ground state by simultaneous two-photon emission, making it more susceptible to the energy splitting.
\begin{figure}[h!]
    \centering
    \includegraphics[width=.85\linewidth]{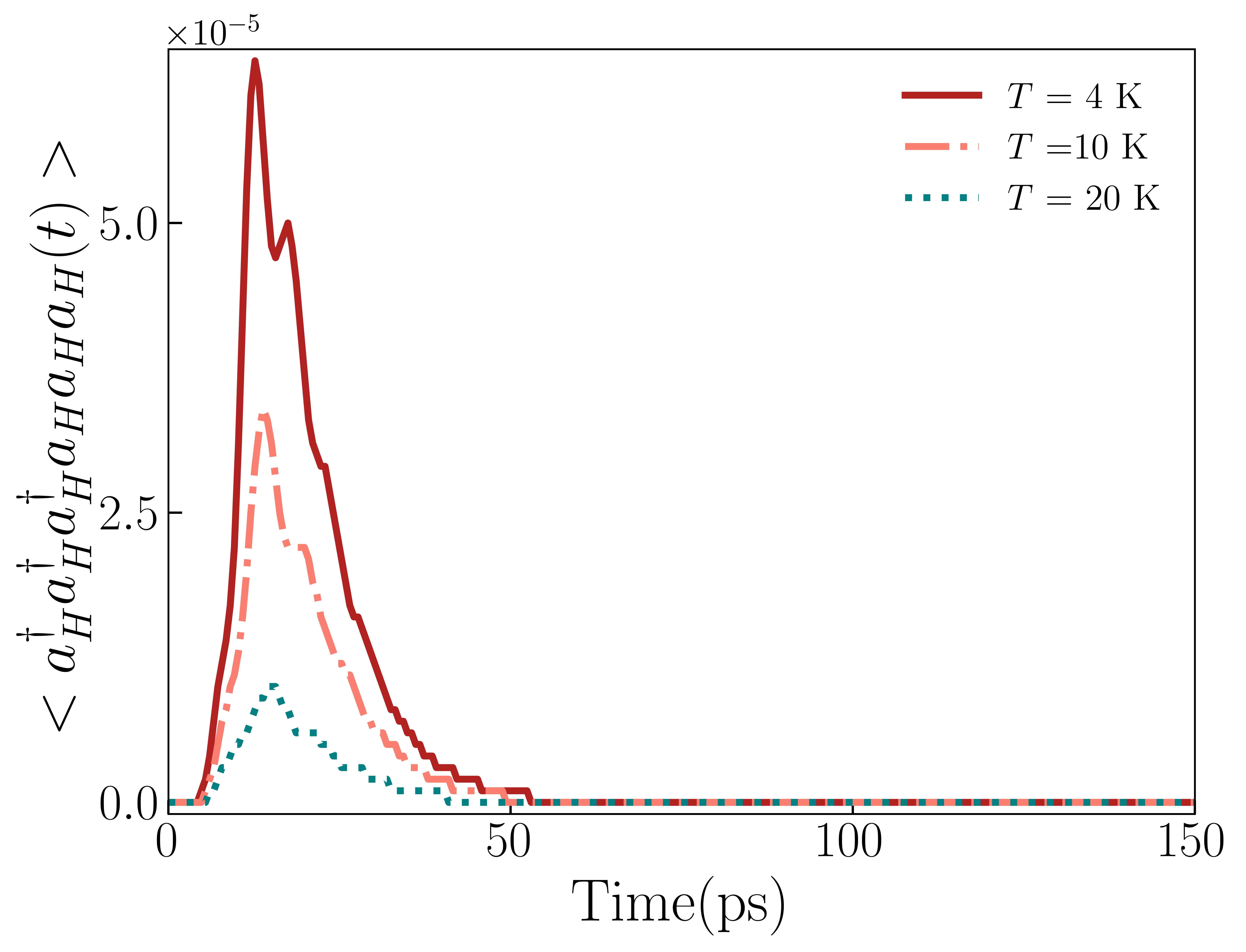}
    \caption{(color online) Temporal dynamics of equal time third order correlation function for different temperatures at $g$ = 70 $\mu$eV.}
    \label{6}
\end{figure}

To inspect the presence of multiphotons, we examine the temporal dynamics of the equal-time third-order correlation function (ETTOCF) $\langle a^{\dagger}_H a^{\dagger}_H a^{\dagger}_H a_H a_H a_H(t)\rangle$. A non-zero value of ETTOCF indicates three or more photons in the system.
The multiphoton emission results from the reabsorption and reemission of cavity photons predominantly by the $H$ polarized exciton state in the presence of $H$ polarized pump pulse. Fig. \ref{6} illustrates that at increased phonon-bath temperatures, the phonon-induced renormalization diminish the effective Rabi frequency and cavity coupling by a factor of $\langle \mathcal{B} \rangle$, consequently inhibiting these processes and resulting in a decreased probability of multiphoton emission. This further signifies the reabsorption and reemission occurrences during the pump pulse.
The emergence of multiphoton emission degrades the coherence between $\vert HH\rangle $ and $\vert VV \rangle$ states and consequently reduces the concurrence.

\begin{figure}[h!]
    \centering
    \includegraphics[width=.85\linewidth]{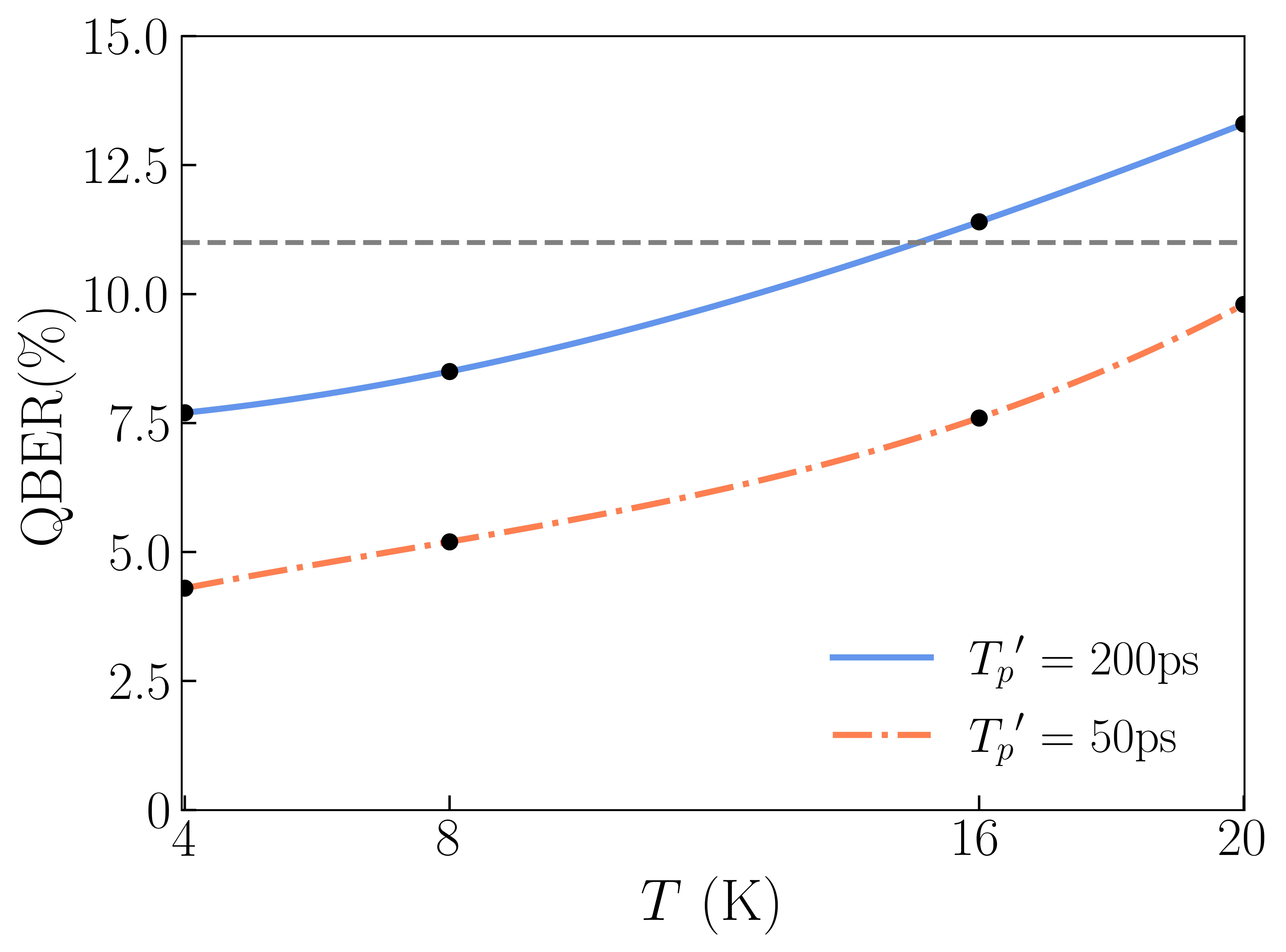}
    \caption{(color online) Qubit error rate as a function of phonon-bath temperature for two different values of $T_{p}^{\prime}$.}
    \label{7}
\end{figure}
Finally, we investigate the effect of exciton-phonon coupling on qubit error rate (QBER) in BBM92 quantum key distribution protocol. The QBER reflects how much the quantum channel has been disturbed by noise or potential eavesdropping. A higher error rate indicates a higher likelihood of eavesdropping or a noisy quantum channel. Here, we assume a perfect quantum channel without any noise or eavesdropping and investigate potential QBER that may arise solely because of phonon-induced degradation of polarization entanglement. In Fig. \ref{7}, we represent the QBER as a function of temperature for $g$ = 70 $\mu$eV. It is clear that the error rate is as low as 7.7 \% at 4 K. However, beyond 15 K, it exceeds the threshold error rate of 11 \%, denoted by the grey dotted line, to detect eavesdropping \cite{Schimpf}. Fig. \ref{7} also indicates that the temporal filtering of phonon-induced decoherence, by reducing the coincidence window time, reduces the QBER. Indeed, we find that the QBER remains below 11\% at ${T_p}^{\prime}$ = 50 ps for temperatures up to 20 K.
\vspace{10pt}
\section{Conclusion}{\label{sec4}}
In conclusion, we have derived a polaron master equation to describe the exciton-phonon interactions in a four-level quantum dot coupled to two orthogonally polarized modes of a pillar microcavity. We have derived the analytical expressions of various phonon-mediated processes and demonstrated that phonon-induced one-photon incoherent rates dominate over others and significantly degrade the entanglement and qubit error rates due to the reduction of coherence between the two-photon states. Furthermore, we have shown that cavity-mediated processes such as cross-coupling between the exciton states and ac-Stark shift of the exciton states contribute to the reduction of coherence only at low temperatures. However, at the elevated temperatures, these detrimental effects get diminished significantly due to phonon renormalization of the cavity coupling strength. It is also shown that the concurrence and qubit error rate could be improved considerably using the temporal filtering of incoherent processes induced decoherence. This work can be employed as a theoretical framework in treating exciton-phonon interaction in a four-level system in the presence of laser drive and cavity coupling. Additionally, the presented results may be useful for experiments on quantum dot-based quantum key distribution with polarization-entangled photons.

\appendix
\setcounter{equation}{0}
\renewcommand{\theequation}{\Alph{section}\arabic{equation}}
\section{ APPENDIX A: Polaron master equation and phonon-mediated relaxation processes}{\label{A}}
Here we provide a detailed analytical description of the polaron master equation (ME). The commutator Eq. \ref{Eq.9} can be expanded and rearranged in terms of system operators. The full polaron ME is given as,
\footnotesize
\begin{align}
        \frac{d\rho}{dt} & \notag=  \frac{1}{i\hbar}[ H^{\prime}_S(t),\rho(t)] +\mathcal{L}_{cav}\rho+\mathcal{L}_{rad}\rho+\mathcal{L}_{deph} \rho \\ \notag   
        &+\Gamma^{+}_H \left \{ \mathcal{L}(a_H \sigma^{\dagger}_{H_1}) + \mathcal{L}(a^{\dagger}_H \sigma_{H_2})  \right \} 
        + \Gamma^{+}_V \left \{ \mathcal{L} (a_V \sigma^{\dagger}_{V_1})+ \mathcal{L}(a^{\dagger}_V\sigma_{V_2}) \right \} \\ \notag
        &+ \Gamma^{-}_H  \left \{ \mathcal{L}(a_H \sigma^{\dagger}_{H_2})
        + \mathcal{L}(a^{\dagger}_H \sigma_{H_1}) \right \}
        + \Gamma^{-}_V  \left \{ \mathcal{L} (a_V \sigma^{\dagger}_{V_2}) 
        + \mathcal{L}(a^{\dagger}_V\sigma_{V_1})  \right \}\\ \notag
        &+\Gamma^{-}_{\Omega}(t)\left \{\mathcal{L}(\sigma^{\dagger}_{H_2})+
        \mathcal{L}(\sigma_{H_1}) \right \}  
        +\Gamma^{+}_{\Omega}(t)\left \{ \mathcal{L}(\sigma^{\dagger}_{H_1})+\mathcal{L}(\sigma_{H_2})\right \} \\ \notag 
        &+ \Gamma^{TP}_{\Omega}(t)\left \{\mathcal{L}(\sigma_{H_1},\sigma^{\dagger}_{H_2})
        +\mathcal{L}(\sigma^{\dagger}_{H_2}, \sigma_{H_1})\right \}\notag \\ 
        &+\Gamma^{I}_{B}\left \{\mathcal{L}(\sigma_{H_1},\sigma^{\dagger}_{H_1}
        \sigma_{H_1})- \mathcal{L}(\sigma^{\dagger}_{H_2},\sigma_{H_2}\sigma^{\dagger}_{H_2})\right \} \notag \\ \notag
        &+\Gamma^{+}_V \left \{\mathcal{L}(a_H \sigma^{\dagger}_{H_1},a_V\sigma^{\dagger}_{V_1})
        + \mathcal{L}(a^{\dagger}_H \sigma_{H_2},a^{\dagger}_V\sigma_{V_2}) \right \} \notag \\
        &+ \Gamma^{+}_H \left \{   \mathcal{L}(a_V \sigma_{V_1}^{\dagger},a_H \sigma_{H_1}^{\dagger})+\mathcal{L}({a_V^{\dagger}\sigma_{V_2},a_H^{\dagger}\sigma_{H_2}}) \right \}  \notag \\
         &- i \Delta^{+}_V \left \{ (a^{\dagger}_H \sigma_{H_1} a_V \sigma^{\dagger}_{V_1}\rho - \rho a^{\dagger}_H \sigma_{H_1} a_V \sigma^{\dagger}_{V_1}) 
        + (a_H \sigma^{\dagger}_{H_2} a^{\dagger}_V \sigma_{V_2} \rho  - \rho a_H \sigma^{\dagger}_{H_2} a^{\dagger}_V \sigma_{V_2} ) \right \} \notag \\
        &- i \Delta^{+}_H \left \{ (a^{\dagger}_V \sigma_{V_1} a_H \sigma^{\dagger}_{H_1}\rho - \rho a^{\dagger}_V \sigma_{V_1} a_H \sigma^{\dagger}_{H_1}) 
        + (a_V \sigma^{\dagger}_{V_2} a^{\dagger}_H \sigma_{H_2} \rho  - \rho a_V \sigma^{\dagger}_{V_2} a^{\dagger}_H \sigma_{H_2} ) \right \} \notag \\
        &+\Big[ \Gamma^{TP}_H \left \{\mathcal{L}(a_H \sigma^{\dagger}_{H_2},a^{\dagger}_H\sigma_{H_1}) \right \}
        + \Gamma^{TP}_V\left \{ \mathcal{L}(a_V \sigma^{\dagger}_{V_2},a_V^{\dagger} \sigma_{V_1})  \right \}  \notag \\
        &+\Gamma^{R}_{B}\left \{ \mathcal{L}(\sigma_{H_1})+ \mathcal{L}(\sigma_{H_2}) 
        -\mathcal{L}(\sigma^{\dagger}_{H_2}, \sigma_{H_1}) \right \}  
        +i \Delta^{p}_{\Omega}  \left \{ (\sigma_{H_2}\sigma_{H_1}\rho - \rho \sigma_{H_2}\sigma_{H_1}) \right \} \notag \\
        &+ i \Delta^{-}_{\Omega}(t) \left \{(\sigma^{\dagger}_{H_2} \sigma_{H_2}\rho - \rho \sigma^{\dagger}_{H_2}\sigma_{H_2} )+(\sigma_{H_1}\sigma^{\dagger}_{H_1}\rho - \rho \sigma_{H_1}\sigma^{\dagger}_{H_1})    \right \} \notag \\ \notag
        &+ \sum_{k = H,V}  i \Delta^{-}_k \left \{  (a^{\dagger}_k \sigma_{k_2} a_k\sigma^{\dagger}_{k_2}\rho 
        - \rho a^{\dagger}_k \sigma_{k_2} a_k \sigma^{\dagger}_{k_2})  
        +(a_k \sigma^{\dagger}_{k_1} a^{\dagger}_k \sigma_{k_1}\rho - 
        \rho a_k \sigma^{\dagger}_{k_1} a^{\dagger}_k \sigma_{k_1}) \right \}  \\
        &+  i \Delta^{-}_{p_k} \left \{ (a^{\dagger}_k \sigma_{k_2} a^{\dagger}_k \sigma_{k_1}\rho - \rho  a^{\dagger}_k \sigma_{k_2} a^{\dagger}_k \sigma_{k_1}) 
         \right \} + H.c. \Big] \label{A1}
\end{align}
\normalsize

The phonon-mediated decay rates and dephasing rates are defined as,
\begin{align}
&\Gamma^{\pm}_H= g^2 \langle \mathcal{B} \rangle^2 \int_{0}^{\infty} d\tau Re \left \{ (e^{\phi(\tau)}-1) e^{\pm i \Delta \tau} \right \} \label{A2} \\
&\Gamma^{\pm}_V= g^2 \langle \mathcal{B} \rangle^2 \int_{0}^{\infty} d\tau Re \left \{ (e^{\phi(\tau)}-1) e^{\pm i (\Delta-\delta) \tau} \right \}\label{A3} \\
&\Gamma^{\pm}_{\Omega}= (\frac{\Omega_H(t)}{2} )^2 \langle \mathcal{B} \rangle^2 \int_{0}^{\infty} d\tau Re \left \{ (e^{\phi(\tau)}-1) e^{\pm i \Delta \tau} \right \} \\
&\Gamma^{TP}_{\Omega}=  (\frac{\Omega_H(t)}{2} )^2 \langle \mathcal{B} \rangle^2 \int_{0}^{\infty} d\tau Re \left \{ (e^{-\phi(\tau)}-1) e^{- i \Delta \tau} \right \} \\
&\Gamma^I_B =  {\Omega_H(t)}^2 \langle \mathcal{B} \rangle^2 \int_{0}^{\infty} d\tau  Re \left  \{ \sinh(\phi(\tau)) \sin(\frac{\Omega^{\prime}_H \tau}{\sqrt(2)}) \right \} \\
&\Delta^{\pm}_H= g^2 \langle \mathcal{B} \rangle^2 \int_{0}^{\infty} d\tau Im \left \{ (e^{\phi(\tau)}-1) e^{\pm i \Delta \tau} \right \} \\
\end{align}
\begin{align}
&\Delta^{\pm}_V= g^2 \langle \mathcal{B} \rangle^2 \int_{0}^{\infty} d\tau Im \left \{ (e^{\phi(\tau)}-1) e^{\pm i \Delta \tau} \right \} \\
&\Gamma^{\textit{TP}}_H= g^2 \langle \mathcal{B} \rangle^2 \int_{0}^{\infty} d\tau Re \left \{ (e^{-\phi(\tau)}-1) e^{- i \Delta \tau} \right \} \\
&\Gamma^{\textit{TP}}_V= g^2 \langle \mathcal{B} \rangle^2 \int_{0}^{\infty} d\tau Re \left \{ (e^{-\phi(\tau)}-1) e^{- i (\Delta-\delta) \tau} \right \} \\ 
&\Gamma^R_B =  2(\frac{\Omega_H(t)}{2} )^2 \langle \mathcal{B} \rangle^2 \int_{0}^{\infty} d\tau  Re \left  \{ \sinh[\phi(\tau)] \big [\cos(\frac{\Omega^{\prime}_H \tau}{\sqrt(2)})-1\big ] \right \} \\
&\Delta^{\pm}_{\Omega}(t)= (\frac{\Omega_H(t)}{2} )^2 \langle \mathcal{B} \rangle^2 \int_{0}^{\infty} d\tau Im \left \{ (e^{\phi(\tau)}-1) e^{\pm i \Delta \tau} \right \} \\
&\Delta^{p}_{\Omega}=  (\frac{\Omega_H(t)}{2} )^2 \langle \mathcal{B} \rangle^2 \int_{0}^{\infty} d\tau Im \left \{ (e^{-\phi(\tau)}-1) e^{- i \Delta \tau} \right \} \\
&\Delta^{-}_{p_H}= g^2 \langle \mathcal{B} \rangle^2 \int_{0}^{\infty} d\tau Im \left \{ (e^{-\phi(\tau)}-1) e^{- i \Delta \tau} \right \} \\
&\Delta^{-}_{p_V}= g^2 \langle \mathcal{B} \rangle^2 \int_{0}^{\infty} d\tau Im \left \{ (e^{-\phi(\tau)}-1) e^{- i (\Delta-\delta) \tau} \right \}  
\end{align}

The polaron master equation, together with the analytic expressions of various phonon-induced processes, allows an insight into different phonon-mediated incoherent processes; for instance,$\Gamma^+_{i = H,V}$ corresponds to the excitation (de-excitation) of the biexciton (exciton) state via cavity photon  absorption (emission), while \(\Gamma^-_{i = H, V}\) represents de-excitation (excitation) of biexciton (exciton) state via cavity photon emission (absorption). The cross-coupling between the exciton states, which is represented by $\left \{\mathcal{L}(a_H \sigma^{\dagger}_{H_1},a_V\sigma^{\dagger}_{V_1})+ \mathcal{L}(a^{\dagger}_H \sigma_{H_2},a^{\dagger}_V\sigma_{V_2}) \right \}$ also depends on $\Gamma^{+}_{i = H,V}$.
Furthermore, $\Gamma_\Omega^\textit{TP}$ denotes the phonon-mediated two-photon processes, wherein the QD is directly transited into the biexcitation (ground) state. In the presence of a finite FSS, $H$ polarized excitons and $V$ polarized excitons encounter different phonon-induced decay rates, as seen from Eq. \ref{A2} and Eq. \ref{A3}. The phonon-assisted incoherent excitation (de-excitation) of biexciton (exciton) states is described by $\Gamma^+_{\Omega}$, while $\Gamma^-_{\Omega}$ represents the phonon-assisted incoherent excitation (de-excitation) of exciton (biexciton) states. The contribution of all other phonon-induced processes is found to be negligibly small. At higher temperatures, the contribution of phonon-induced processes due to the cavity effects is found to be much smaller than laser-induced decay and dephasing rates.

\section{APPENDIX B: Details on the phonon-modified system operators}
{\label{B}}
\setcounter{equation}{0}
To treat exciton-phonon coupling nonperturbatively, we carry out a unitary polaron transformation
$H^{\prime} = e^P H e^{-P}$
to diagonalize the electron-phonon coupling part of the Hamiltonian. The polaron frame transformed Hamiltonian is, $H^{\prime} = H_S^{\prime}+ H_B^{\prime}+H_I^{\prime}$.
To expand the two-time phonon system operators in terms of the one-time operators in the interaction pictures, we use the polaron-transformed system Hamiltonian $H^{\prime}_S$.  The transformation is given by, $X_m(t,\tau ) = e^{-iH^{\prime}_S\tau/\hslash}X_m e^{iH^{\prime}_S\tau/\hslash} $. Here, we have considered the dot-cavity coupling $g$ is much smaller than dot-cavity detuning $\Delta$ and Rabi frequency associated with the horizontally polarized pulse, i.e, $g \ll \Delta$ and $g \ll \Omega_H$. Under these conditions, $H^{\prime}_S$ reduces to,
 $H_S^{\prime}= \hslash \Delta \vert {H}\rangle  \langle {H}\vert + 
 \hslash \Delta \vert {V}\rangle  \langle {V}\vert +  \frac{\hslash \Omega^{\prime}_H(t)}{2}(\sigma^{\dagger}_{H_1}+\sigma^{\dagger}_{H_2}+ \sigma_{H_1}+\sigma_{H_2})  $, where, $ \Omega^{\prime}_H(t) = \langle \mathcal{B} \rangle \Omega_H(t)$.
Then, using the unitary transformation, we may derive,
{\small
\begin{align}
  X_g (t,\tau) = \notag & \hslash g \big [ (a^\dagger_H \sigma_{H_1}+ a_H \sigma^{\dagger}_{H_2} ) e^{-i \tau \Delta} +(a^\dagger_H \sigma_{H_2}+a_H \sigma^{\dagger}_{H_1} ) e^{i \tau \Delta}\big ] \\ \notag
  &+ \hbar g \big [(a^\dagger_V \sigma_{V_1} + a_V\sigma^{\dagger}_{V_2})e^{-i (\Delta-\delta)\tau}\\ \notag
  &+( a^\dagger_V \sigma_{V_2} +a_V\sigma^{\dagger}_{V_1}) e^{i (\Delta-\delta)\tau} \big ]\\ \notag
  &+ \frac{\hslash \Omega_H(t)}{2}
  [(\sigma_{H_1}+\sigma^{\dagger}_{H_2}) e^{-i \tau \Delta}+(\sigma_{H_2}+\sigma^{\dagger}_{H_1}) e^{i \tau \Delta}] \\
\end{align}

\begin{align}
  X_u (t,\tau) \notag& =  i\hslash g \big [ (a^\dagger_H \sigma_{H_1} 
 -a_H \sigma^{\dagger}_{H_2} ) e^{-i \tau \Delta}+ (a^\dagger_H \sigma_{H_2} -a_H \sigma^{\dagger}_{H_1} ) e^{i \tau \Delta} \big ] \\ \notag
     &+i \hslash g \big [
    (a^\dagger_V \sigma_{V_1}- a_V\sigma^{\dagger}_{V_2})e^{-i (\Delta-\delta)\tau}  +(a^\dagger_V \sigma_{V_2} 
    -a_V\sigma^{\dagger}_{V_1} )e^{i (\Delta-\delta)\tau}  \big ]  \\ \notag
    &+i \frac{\hslash \Omega_H(t)}{2}
    [(\sigma_{H_1}-\sigma^{\dagger}_{H_2}) e^{-i \tau \Delta}+(\sigma_{H_2}-\sigma^{\dagger}_{H_1}) e^{i \tau \Delta}] \\ \notag
    &+i\hslash \Omega_H(t) \sin({\frac{\Omega^{\prime}_H \tau}{\sqrt{2}}}) (\sigma^{\dagger}_{H1} \sigma_{H_1}-\sigma_{H_2} \sigma^{\dagger}_{H2} )  \\ \notag
    &+i\frac{\hslash \Omega_H(t) }{2}[\cos({\frac{\Omega^{\prime}_H \tau}{\sqrt{2}}})-1](\sigma_{H_1}-\sigma^{\dagger}_{H_2}+\sigma_{H_2}
   -\sigma^{\dagger}_{H_1}) \\
\end{align}
}
For a QD-driven system, we only include terms proportional to $g^2$ and $\Omega_{H}^2$ and exclude cross terms proportional to $g \Omega_H$ for preserving the Lindblad form; also, such terms do not contribute significantly in the considered parameter regime \cite{Hughes}.

\section*{Acknowledgment}

U.D. gratefully acknowledges a research fellowship from MoE, Government of India. P.K. acknowledges grant from Anusandhan National Research Foundation (ANRF), Government of India (Grant No. SRG/2023/000560). A.K.S. acknowledges the grant from MoE, Government of India (Grant No. MoESTARS/STARS-2/2023-0161).

\subsection*{Disclosures}
The authors declare no conflicts of interest.

\subsection*{Data availability}
The data that support the findings of this study are available upon reasonable request from the authors.

\end{document}